\documentstyle[prl,aps,twocolumn]{revtex}

\begin{document}

\title{Controlling spatiotemporal dynamics with time-delay feedback}
\author{M.~E.~Bleich and J.~E.~S.~Socolar}
\address{Department of Physics and Center for Nonlinear and Complex Systems,
	 Duke University, Durham, NC 27708}
\date{\today}

\maketitle

\begin{abstract}
We suggest a spatially local feedback mechanism for stabilizing
periodic orbits in spatially extended systems.
Our method, which is based on a comparison between present and
past states of the system, does not require the external generation
of an ideal reference state and can suppress both absolute and
convective instabilities.
As an example, we analyze the complex Ginzburg-Landau equation
in one dimension, showing how the time-delay feedback enlarges
the stability domain for travelling waves.
\end{abstract}


A common situation encountered
in the operation of physical systems or devices is that
a useful solution of the equations of motion turns out
to be unstable in a parameter regime of interest.
In many cases the desired behavior is
a steady state or a regular periodic motion, and the
instability eventually leads to chaotic fluctuations
which limit the system's performance.
Thus one is led to explore the possible modifications
of the system that render the desired motion stable.
 
Recently, there has been intense interest in the application
of proportional feedback for stabilizing periodic orbits.
Since the orbit in question is a solution to the equations
of motion, the stabilizing feedback signal vanishes
when control is successful, so that all the
desirable features of the uncontrolled system are retained.
Many methods for controlling systems with only a few
relevant degrees of freedom have 
now been successfully demonstrated. \cite{ccreview,ETDAS}

Some of the most interesting and significant dynamical instabilities
arise in spatially extended systems
which may be described by partial differential equations,
a large number of coupled ordinary differential equations,
or coupled map lattices.
Well-known examples of practical interest include convecting fluids,
large Fresnel number lasers, and arrays of semiconductor lasers.
For small systems, the number of unstable modes remains small and
techniques involving only a few degrees of freedom can effectively
treat the spatiotemporal dynamics. \cite{fewmodes}
For the case of open systems with convective instabilities, control of larger
systems has also been demonstrated. \cite{convective}

In this paper, we present and analyze a new method for stabilizing
periodic orbits in arbitrarily large systems. 
(A different approach has been suggested by Hu and Qu. \cite{lattice})
Our approach is a generalization of 
the technique known as ``extended time-delay autosynchronization'' (ETDAS),
which has been successfully applied to a variety of low-dimensional systems,
both numerical and experimental. \cite{ETDAS}
In ETDAS, the current state of the system is compared to
its state one or more periods in the past and a feedback
signal is generated locally (at each spatial point) based
on the local value of the difference.
Our analytical treatment of the important special case of the 
1D complex Ginzburg-Landau equation shows both that this method can
stabilize spatially extended periodic orbits
and, more generally, that the introduction of spatially local
time-delayed interactions can dramatically alter the
stability properties of extended deterministic systems.

Our method has three key features:
First, it applies equally well to systems with
absolute or convective instabilities.
Second, it is, by construction, scalable up to arbitrarily
large system sizes with no increase in complexity.
Third, it does not require comparison to an external
reference signal and therefore might be
implemented in fast (optical) systems,
systems in which the reference state is not known a priori,
or systems in which the reference state has 
nontrivial spatiotemporal structure.
In systems where the control works, the only information
that must be supplied by the controller is the period $\tau$
of the desired motion.

The general approach we take is as follows.
To generate the feedback signal for a system described
by an evolving field $\phi(x,t)$,
the entire field is compared to time-delayed
images of itself $\phi(x,t-n\tau)$, with $\tau$
chosen to be the period of the desired orbit
and $n$ taking all positive integer values.
With $t_n\equiv t-n\tau$, 
the feedback signal is the field 
\begin{eqnarray}
\label{phi}
 \epsilon_\phi (x,t) = 
 \gamma \sum_{n=0}^\infty R^{n}\left( \phi(x,t_n)- \phi(x,t_{n+1}) \right),
\end{eqnarray}
where $\gamma$ is a real parameter (the gain) and
$R$ is a real parameter between 0 and 1.
We assume here that $\tau$ is known in advance.
(The problem of finding appropriate periodic orbits and their periods
is beyond the scope of this paper.)
It is clear that $\epsilon_\phi$ 
vanishes identically when $\phi(x,t)$ is periodic
in time with period $\tau$.
As emphasized elsewhere \cite{ETDAS}, the infinite sum
can be obtained in practice with a recursive feedback loop
that contains only a single time-delay device. 
In some systems, it may be possible to implement
this form of control directly; e.g., by using optical
elements that preserve the spatial structure of a laser beam.
Alternatively, $\epsilon_\phi (x,t)$ can be considered as
a limiting case of the placement of independent
ETDAS controllers at many discrete 
points in the system.

We are interested in the extent to which
proper choices of $\gamma$ and $R$ can improve the
stability of selected time-periodic patterns
of the field $\phi(x,t)$ for arbitrarily large system sizes.
In this paper, we treat the 
1D complex Ginzburg-Landau (CGL) equation with a cubic nonlinearity,
a partial differential equation that describes 
a large class of systems that undergo
a bifurcation from regular oscillations to spatiotemporal chaos. \cite{cgl}
In addition to its relevance to fluid and laser systems,
this equation has the advantage of possessing purely
sinusoidal travelling wave solutions which permit
a detailed analysis.
Our linear stability analysis shows that periodic
travelling waves states can indeed be stabilized
by appropriate choices of $\gamma$ and $R$,
even for system parameters corresponding to
turbulence (``defect chaos'' \cite{cgl}) in the uncontrolled equation.
An numerical illustration of successful control is shown in Fig.~1.  

The controlled CGL equation we study may be written in dimensionless form as
\begin{eqnarray}
\label{cgl}
\partial_t A = A + (1+ i c_1) \partial_x^2 A
	- (1- i c_3) \left| A \right|^2 A + \epsilon_A,
\end{eqnarray}
where $x$ is a one-dimensional continuous variable,
$A(x,t)$ is a complex field, 
$c_1$ and $c_3$ are real
parameters, and $\epsilon_A(x,t)$ is the control term defined above.
Without the control term, 
Eqn.~(\ref{cgl}) admits travelling wave solutions of 
wavenumber $k$ and frequency $\omega = -c_3 + (c_1+c_3)k^2$.
Each solution
\begin{equation}
A_k(x,t) = \sqrt{1-k^2}\;\exp(ikx-i\omega t)
\end{equation}
becomes unstable for large enough
$c_1$ and/or $c_3$, and all of them
are unstable for $c_1 c_3 > 1$. \cite{newell}

We find that 
when $\tau$ is chosen to be $2\pi/\omega$,
the domain of $c_1$ and $c_3$ values over which
the solution $A_k$ is stable is expanded significantly
for modest values of $\gamma$ and $R$.
Some typical results are shown in Fig.~2.
Each panel of the figure corresponds to a different
set of values for $k$ and $R$.
The shaded region represents the parameter values for which the
travelling wave $A_k$ is a stable solution of the 
equation with control, but would be unstable without control.
The instability would be convective just above onset,
but absolute for larger $c_1$ or $c_3$ \cite{abscon},
as indicated by the dashed line in Fig.~2.
A surprising result is that time-delay feedback
allows stabilization of travelling waves deep
into the ordinarily chaotic, absolutely unstable regime 
($c_1 c_3 \sim 8$),
even though it is almost totally ineffective
in stabilizing the uniform ($k=0$) oscillatory state.

We now describe our procedure for obtaining the stability domains
depicted in Fig.~2.
Standard linearization of Eq.~(\ref{cgl})
about $A_k(x,t)$ yields sets of ordinary, time-delay differential
equations for the Fourier amplitudes of a perturbation.
The technique of Ref.~\cite{BS} is then applied to determine
the stability of the different modes.
In each period $\tau$, a given mode grows or decays by a complex factor
$\mu$ (a Floquet multiplier).
A system is stable if and only if $\left| \mu \right| < 1$ for every mode.
The defining relation for the Floquet multipliers of a general,
finite-dimensional system controlled by ETDAS is, \cite{BS}  
\begin{equation}
\label{condition}
\left| \mu^{-1} T \left[ e^{\int_0^\tau dt \left( {\bf J} + 
\gamma \frac{1-\mu^{-1}}{1-R \mu^{-1}} {\bf M} \right)} \right] - 
1\!\!1 \right| = 0,
\end{equation}
where $T[\ldots]$ represents the time-ordered product,
$\bf{J}$ is the Jacobian of the uncontrolled mode equations, ${\bf M}$ 
is a ``control matrix'' that contains the information about the way in
which the control signal is formed and
enters into the dynamical equations, and $1\!\!1$ is the identity matrix.

In the present case, three features simplify the analysis:
first, ${\bf M} = 1\!\!1$;
second, the Fourier modes decouple, with each
yielding a condition of the form of Eqn.~(\ref{condition}) 
with $2\times 2$ matrices ${\bf J}$ and ${\bf M}$;
third, neither ${\bf J}$ nor ${\bf M}$ is time dependent.
The latter is due first to the 
trivial time-dependence of the desired solution $A_k$
and second to the directly additive way in which
$\epsilon$ appears in the equation for the controlled system.
In such cases,
the determinant in Eqn.~(\ref{condition}) may be evaluated explicitly
by solving the differential equation that yields the time-ordered product.
Here the defining relation for the Floquet multipliers 
associated with a perturbation at wavenumber $q+k$ becomes,
\begin{eqnarray}
\label{simpcond}
g(\mu^{-1}) & \equiv & \mu^{-2} e^{2 \alpha \tau} - 
2 \mu^{-1} e^{\alpha \tau}
 \cosh{(\beta \tau)} + 1 = 0 \; ; \\
 \alpha & = &  - q^2 - 2 i c_1 k q + k^2 -1 + 
 \gamma \frac{1-\mu^{-1}}{1-R \mu^{-1}},\nonumber \\
\beta & = &  [\left(1-k^2 - 4 i c_3 k q + 2 c_1 c_3 q^2\right)
 \left(1-k^2\right) \nonumber \\
 &\ & + 4 k^2 q^2 + 4 i c_1 k q^3 - c_1^2 q^4]^{\frac{1}{2}}. \nonumber
\end{eqnarray}

The state $A_k(x,t)$ with particular 
choices of $c_1$, $c_3$, $\gamma$, and $R$
is linearly stable if and only if all of the
roots of $g$ lie outside the unit circle.
Note that $g(\mu^{-1})$ has an infinite number of roots
due to the time delay in the system.
As described in Ref.~\cite{BS}, it is straightforward
to perform a winding number calculation 
(or a contour integration) that will return ${\cal N}$,
the number of roots that lie inside the unit circle.
The linear stability condition then reduces to ${\cal N} = 0$.
The winding number calculation is performed numerically
by evaluating $g$ on selected points on the unit circle.  
The precision of such a technique is determined
by how well one can distinguish between a root inside the unit circle
and one very close, but still outside.  
Using an adaptive step-size method, we resolved the location
of such roots to an accuracy of $10^{-6}$.

It is at this point that the many degrees of freedom 
in a spatially extended system complicate the analysis.
In order for a particular state to be stable,
it must be true that a single value of $\gamma$ exists
for which $A_k$ is stable with respect to 
perturbations at all wavenumbers.
To see whether such a $\gamma$ exists for fixed $k$, $c_1$, $c_3$, and $R$,
it is useful to plot the region of stability in the space
of $\gamma$ and the perturbation wavenumber $q$. \cite{crawl}
Fig.~3(a) shows an example for which 
$A_k$ is linearly
stable against perturbations of all wavenumbers for a range of $\gamma$ 
(shown between the dashed lines). 
Note that the plot must be symmetric about $q=0$ since
from Eqn.~\ref{simpcond}
it is clear that $\mu(q) = \mu^*(-q)$.
The rapid divergence of the stability boundaries
for large $q$ merely reflects  
the fact that the system is highly stable with respect to
large $q$ perturbations in the absence of control.

Figs.~3(b) and (c) show why control cannot
be achieved for some values of $c_1$, $c_3$, $k$, and $R$.
The problem is that peaks in the lower boundary reach to values
of $\gamma$ that are already ruled out by valleys in the upper boundary,
so that no single value of $\gamma$ can stabilize all wavenumbers.
The source of the peaks may be understood as follows:
For a periodic state with frequency $\omega$
and ${\bf J}$ and ${\bf M}$ independent
of time, it can be shown that
ETDAS cannot stabilize a perturbation 
for which $\arg\;\mu=m\omega$, where $m$ is any integer.
The peaks in the lower boundary occur at
wavenumbers where this condition is approximately satisfied.
In Fig.~3, the peak in the lower boundary at $q=0$ 
corresponds to $m = 0$.  
In Fig.~3(c), the peak at $q\sim 0.75$ corresponds to $m = \pm 1$.
In the present case there can be no effect from higher $\left| m \right|$
because control will already have been lost due to $m = \pm 1$.

By analyzing stability diagrams in the $q-\gamma$ plane 
for a grid of values in the $c_1-c_3$ plane, 
one can construct the stability diagrams
shown in Fig.~2.
As in the example of Fig.~3(a), there is a range of the
feedback gain $\gamma$
that successfully achieve control for each point in
the shaded area of Fig.~2.  
In general, the minimum value of $\left| \gamma \right|$
required increases smoothly as $c_1$ and/or $c_3$ increases.
In the domains shown here,
as well as others we investigated,
the value of $\left| \gamma \right|$ required
for stability is less than $1$ 
even at the highest values of
$c_1$ and $c_3$ in the controllable domain.
(Details will be given in a longer paper.)
Beyond the line labelled U0 (U1) control is lost due to the 
mechanism described above with $m=0$ ($m=\pm 1$), 
{\em not} through a divergence in the required $\gamma$.

Fig.~2(a), (c), and (d) illustrate
how the stability boundaries shift for different
choices of $k$.
Fig.~2(a) corresponds to $k=.075 \pi$.  
For larger $k$ (Fig.~2(c)), 
the U0 line moves farther from
the uncontrolled stability boundary
and U1 moves closer.
For smaller $k$ (Fig.~2(d)),
the situation is reversed. 
As $k$ is decreased toward $0$, the boundary U0 approaches
the original uncontrolled stability boundary, so that
no enhacement of the $k=0$ state is obtainable.
Figs.~2(a) and (b) show the effect of changing $R$.
As $R$ is increased, the 
domain of stability increases in area.
However, even as $R$ approaches its maximum value of 1, 
the domain of stability 
cannot include the region in which one of the unstable modes of the
uncontrolled system has frequency $\pm \omega$.
The boundary of this region is the dotted line in 
Fig.~2(a).

We have checked specific aspects of the results 
presented in Fig.~2
with numerical simulations of the controlled CGL equation.
Periodic boundary conditions were employed with the 
system size chosen to be an integer multiple of
the wavelength of the travelling wave.
System sizes corresponded to 
a length of at least $15\times 2\pi/k$.
The simulations were performed with a second-order 
predictor-corrector and finite difference technique. 
with time steps of order $10^{-2}$ and spatial resolution 
$\sim 400$ points.
The instabilities were observed to occur at 
values of $(c_1,c_3)$ consistent
with the analytic results presented here for infinite systems.

We have demonstrated that 
time-delay feedback can be effective in stabilizing 
periodic states of spatially extended systems.
Application of this technique to the stabilization
of unstable ordered states in fluid, laser, and biological
systems is strongly suggested.
We expect the control technique to be applicable
to many types of periodic states, though the stability analysis
may become complicated.
If the linearized equation for the
perturbations about the periodic solution has space-dependent coefficients
but its time dependence is still trivial,
perturbations can be decomposed into appropriate eigenfunctions 
and the analysis discussed here will apply.
If the periodic state has trivial spatial dependence
but nontrivial time dependence, then the stability of the Fourier modes
can be analyzed using the numerical method of Ref.~\cite{BS}.
Finally, when the periodic state has complicated spatiotemporal structure,
it appears that numerical integration 
of the controlled equations would be the
most efficient approach.
Even in the absence of any stability analysis, however, control can
be attempted in a physical system given only the 
knowledge of $\tau$ and the ability to 
adjust the single parameter $\gamma$.

Our work points to several important questions for future study.
What is the minimum density of discrete controllers needed
in situations where spatially continuous processing in the
feedback loop is not possible?
What level of noise can be tolerated?
How can one force the system from the spatiotemporally chaotic state
into the desired controllable state?

Finally, we suggest that the application of time-delayed feedback
may be a valuable tool for studying the intrinsic physics of 
spatiotemporally chaotic systems.
By varying $\gamma$ and $\tau$ slowly, it may be possible to
locate previously unknown periodic states or 
to observe other novel effects.

We thank D. Gauthier and  H. Greenside for 
useful conversations and critical readings of the manuscript.
JESS gratefully acknowledges the hospitality of the Aspen Center for Physics, 
where some of this work was done.  
The work was supported by NSF grant DMR-94-12416.


\begin{references}
\bibitem{crawl} The stability boundaries are found numerically, 
		as in Ref.~\cite{BS}.
\end{references}
\end{document}